\documentclass[11pt,twoside]{article}
\usepackage{newpasp}

\index{Sellwood, J. A.}
\index{Kosowsky, A.}

\def\eg{{\it e.g.}}
\def\etal{{\it et al.}}

\def\ie{{\it i.e.}}

\begin{document}

\title{Does Dark Matter Exist?}
\author{J. A. Sellwood and A. Kosowsky}
\affil{Rutgers University, Department of Physics \& Astronomy, \\
136 Frelinghuysen Road, Piscataway NJ 08854-8019, \\
sellwood/kosowsky@physics.rutgers.edu}

\begin{abstract}
The success of the $\Lambda$CDM model on large scales does not extend down to 
galaxy scales.  We list a dozen problems of the dark matter hypothesis, some of 
which arise in specific models for the formation of structure in the universe, 
while others are generic and require fine tuning in any dark matter theory.  
Modifications to the theory, such as adding properties to the DM particles 
beyond gravitational interactions, or simply a better understanding of the 
physics of galaxy formation, may resolve some problems, but a number of 
conspiracies and correlations are unlikely to yield to this approach.  The 
alternative is that mass discrepancies result from of a non-Newtonian law of 
gravity, a hypothesis which avoids many of the more intractable problems of dark 
matter.  A modified law of gravity is not without formidable difficulties of its 
own, but it is no longer obvious that they are any more daunting than those 
facing DM.
\end{abstract}


\section{Introduction}
Discrepancies between the apparent baryon content of galaxies and galaxy 
clusters and their dynamically inferred masses are generally attributed to dark 
matter (DM).  A still larger mass discrepancy exists out to the horizon scale, 
since Big Bang nucleosynthesis requires a low baryonic mass fraction in the 
universe (Tytler \etal\ 2000), which nevertheless appears to have flat geometry 
(de Bernardis \etal\ 2000; Hanany \etal\ 2000).  The missing energy density 
required by conventional cosmology is now supposed to reside not only in 
non-baryonic particles, but also partly as vacuum energy.

The currently popular $\Lambda$CDM model for the evolution of structure in the 
universe is quite successful on large scales (\eg\ Bahcall \etal\ 1999), but its 
predictions on galaxy scales disagree with observed properties of galaxies.  A 
number of lines of evidence suggest that the problem is more fundamental than 
simply our inadequate understanding of the process of galaxy formation.

\section{Mass Distributions in the Inner Parts of Galaxies}
Before listing the problems of DM, we briefly review two properties of the mass 
distribution within galaxies that exacerbate some of the difficulties.  Our 
conclusions are controversial, but we find strong evidence for both points.

First, putative galaxy halos seem to have cores of finite, as opposed to cusped, 
central densities.  Density estimates from rotation curves of low-luminosity and 
low surface brightness (SB) galaxies are most revealing, since these galaxies 
have the largest mass discrepancies and therefore halos for which compression by 
baryonic infall is least important.  Van den Bosch \etal\ (2000) stress that the 
inner slope in HI data can be underestimated because of ``beam smearing,'' and 
Swaters \etal\ (2000) indeed find that optical data often indicate a steeper 
rise; but their cases either require, or are consistent with, a finite density 
core -- whatever M/L is ascribed to the luminous component.  Other well-resolved 
rotation curves of low-luminosity galaxies (Rubin \etal\ 1985; Courteau 1997; 
Sofue \etal\ 1999; Blais-Ouellete \etal\ 2000a,b; Palunas \& Williams 2000) 
generally also reveal a gentle rise, indicating that their halos do indeed have 
low-density cores.

Second, the mass in the inner parts of bright galaxies is dominated by the 
luminous disk and bulge.  ``Maximum disk'' models are strongly suggested by the 
impressive match between the shapes of inner rotation curves and predictions 
from the luminous matter alone (Kalnajs 1983; Kent 1986; Broeils \& Courteau 
1997; Palunas \& Williams 2000), and other evidence is reviewed by Bosma (1999) 
and Freeman (this meeting).  Furthermore, recent work on barred galaxies 
(Debattista \& Sellwood 1998, 2000; Weiner \etal\ 2000) has shown that any DM 
halo makes a negligible contribution to the inner rotation curve even after the 
formation of the disk and bulge, and that this conclusion for barred galaxies 
also holds for their unbarred counterparts.

\section{Problems Created by the Dark Matter Hypothesis}
The problems we list here mostly relate to properties of galaxies.  The first 
ten are well established and are ordered such that those which are generic to 
any form of DM come before those specific to particular DM models.  The last two 
are less well established observationally but are potentially just as thorny.  
The physics of galaxy formation is undeniably messy, creating scope for 
counter-arguments that may avert individual difficulties.  We do not have space 
here to evaluate the many such arguments that have been advanced, but no 
combination of such ideas comes anywhere near to resolving all these problems.

\begin{enumerate}

\item The ``disk-halo conspiracy'' (Bahcall \& Casertano 1986) describes the 
absence of a feature in galaxy rotation curves at which the dominant source of 
central attraction changes from luminous matter to dark.  Many galaxies are now 
known in which the rotation curve does drop somewhat at the edge of the visible 
disk (\eg\ Casertano \& van Gorkom 1991; Verheijen 1997; Bosma 1999), but it is 
extremely rare for the drop to exceed about 10\%.  Blumenthal \etal\ (1986) 
showed that a featureless rotation curve is expected if DM dominates galaxies 
right to their centers, but it is much harder to understand why the circular 
orbital speed from the luminous matter, which dominates the inner region (see 
\S2), should be so similar to that from the DM at larger radii.  For any galaxy 
dominated by stars in its center, initial conditions for the dark and luminous 
matter must be finely tuned to produce a flat rotation curve.

\item Extreme low-SB galaxies lie on the same Tully-Fisher relation (TFR) 
derived from high-SB galaxies, with somewhat greater scatter (Zwaan \etal\ 1995; 
Sprayberry \etal\ 1995; McGaugh \etal\ 2000).  Thus we observe similar circular 
speeds in all galaxies of a given luminosity, no matter how widely the luminous 
material is spread.  This amazing result requires that the overall M/L of the 
galaxy rises with decreasing SB in just the right way so as to preserve a tight 
relation between total luminosity and circular speed.  Either the true M/L of 
the stellar population changes with surface brightness, which seems unlikely (de 
Blok \& McGaugh 1997), or the DM fraction rises as the luminous surface density 
declines.  The needed variations would be minor if DM dominated in all galaxies, 
but since stars dominate the mass in the inner parts of high-SB galaxies (\S2), 
eliminating the SB dependence again requires careful tuning.

\item Sanders (1990), Milgrom (1998) and McGaugh (1999a) show that mass 
discrepancies begin to be detectable only when the acceleration drops below 
$\sim 10^{-8} \hbox{ cm s}^{-2}$.  Any DM model must reproduce this 
characteristic acceleration scale over a wide range of galaxy sizes, but none 
has yet done so convincingly.

\item Aside from the disk M/L, DM halo fits to rotation curves generally employ 
two extra parameters: \eg\ the core radius and asymptotic velocity, or the scale 
radius and concentration index.  Actual galaxy rotation curves do not require 
all this freedom, however, since they can be fitted with only the disk M/L as a 
parameter (\eg\ Sanders \& Verheijen 1998) when a modified gravitational force 
law of the MOND form (Milgrom 1983) is employed.  Any DM model must therefore 
contain a physical mechanism that relates the halo parameters to the luminous 
mass distribution.

\item A merging hierarchy causes the cooled baryonic fraction to lose angular 
momentum to the halo, making disks that are too small (Navarro \& White 1994; 
Navarro \& Steinmetz 1997).  The predicted angular momentum of the disk is at 
least an order of magnitude less than that observed.  The problem is only 
partially ameliorated (MacLow \& Ferrara 1999; Navarro \& Steinmetz 2000b) if 
some process (usually described as ``feedback from star formation'') prevents 
most of the gas from cooling until after the galaxy is assembled.  While this 
difficulty is best known within the CDM context, merging protogalaxies in any 
hierarchical structure formation model with DM will involve chronic angular 
momentum loss to the halos (\eg\ Barnes 1992).

\item Every collapsed halo should manifest the same peak phase space density, 
$f_{\rm max}$, if DM is collisionless, was initially homogeneously distributed, 
and had an initially finite $f_{\rm max}$ (Liouville's theorem).  Further, the 
finite central density of galaxy halos (\S2) both suggests that halos collapsed 
from material having in initially finite $f_{\rm max}$ (infinite initial phase 
space density forms cusped halos) and also allows $f_{\rm max}$ to be estimated 
easily.  The spectacular variation of $f_{\rm max}$ between galaxies found by 
Sellwood (2000) and Dalcanton \& Hogan (2000) indicates that DM cannot be a 
simple collisionless particle.

\item High-resolution simulations that follow the formation and evolution of 
individual galaxy halos in CDM find strongly cusped density profiles (Moore 
\etal\ 1998; Klypin \etal\ 2000) even before the baryonic component cools and 
settles to the center.  No observational evidence {\it requires\/} halos to have 
the predicted cusps.  Further, the ``concentration index'' has a wide range 
(Bullock \etal\ 1999), but most fits to rotation curves yield values well below 
the predicted range in all types of galaxy (\S2) -- even the Milky Way (Navarro 
\& Steinmetz 2000a).

\item Navarro \& Steinmetz (2000a) describe their failure to predict the 
zero-point of the TFR as a ``fatal problem for the $\Lambda$CDM paradigm.''  
They show that no matter what M/L is assumed for the disk, the predicted 
circular speed at a given luminosity is too high because the halo density is too 
high.

\item Simulations produce numerous sub-clumps within large DM halos (Klypin 
\etal\ 1999; Moore \etal\ 1999).  The clumps are more numerous than the numbers 
of observed satellite galaxies, and may threaten the survival of a thin disk in 
the host galaxy.

\item The TFR discrepancy is even worse, since CDM predicts $L \propto V^3$ 
(Dalcanton, Spergel \& Summers 1997; Mo, Mao \& White 1998), whereas Verheijen 
(1997) stresses that when $V$ is interpreted as the circular velocity of the 
flat part of the rotation curve, the true relation is very nearly $L \propto 
V^4$.  Any mechanism which systematically boosts luminosity as a function of 
mass must also reproduce the very small scatter in the TFR.

\item The DM halos that form in simulations are generally tri-axial (Dubinski \& 
Carlberg 1991; Warren \etal\ 1992), but become nearly oblate in their inner 
parts when a disk is added (Dubinski 1994).  Current constraints on halo shapes 
(Sackett 1999) are generally thought to be consistent with these predictions.  
However, the halo of NGC 2403 seems to become more nearly axisymmetric at larger 
radii (Schoenmakers 1998), opposite to the CDM expectation, and Franx \etal\ 
(1994) find IC 2006 to be impressively round at $6 R_e$.  Much more data are 
needed to determine whether this behavior is typical or anomalous.

\item The first precision measurements of the microwave background power 
spectrum at sub-degree scales show a second acoustic peak greatly suppressed 
compared to the first (de Bernardis \etal\ 2000; Hanany \etal\ 2000), producing 
an uncomfortable fit to standard cosmological models (\eg\ Lange \etal\ 2000; 
Tegmark \etal\ 2000).  McGaugh (1999b) has pointed out that unforced acoustic 
oscillations, as might be expected in the absence of dark matter potential 
wells, give such a peak height ratio.

\end{enumerate}

\noindent
Only three of these problems hinge on the properties mentioned in \S2: If disks 
in bright galaxies are significantly sub-maximal, problem 1 would largely go 
away and 7 would be weakened (while 2 would be altered, not solved).  If halos 
have mild density cusps, problem 6 would go away and 7 would again be weakened.

\section{Discussion}
The first five problems on our list are the most generic and intractable in DM 
models.  Problems 1, 2 \& 4 do not merely arise from mismatches between theory 
and observation, but require fine tuning in any DM scenario.  Any physical 
process to account for them must somehow couple the dark and luminous 
components.  If the only interaction between the baryons and DM is 
gravitational, it is hard to imagine how such a mechanism could arise.

Problems 7--10 are well-rehearsed difficulties of CDM.  The possibility that 7 
\& 8 could be solved by a more realistic treatment of baryonic processes alone 
is remote since they relate to the total mass distribution which, in the 
simulations, is dominated by DM.  Even if the baryonic mass fraction is 
significant, it is unclear what process could loosen tight DM concentrations; 
for example, Debattista \& Sellwood (2000) show that an intolerable degree of 
dynamical friction has a very mild effect on the halo density profile.  Problems 
9 \& 10 could conceivably be solved by some kind of complex baryonic physics 
(\ie\ star formation), but this solution to 10 would require more fine tuning.

Another approach to problems 7--10 is to modify the assumed properties of the DM 
particles.  Warm DM (\eg\ Colombi, Dodelson \& Widrow 1996; Sommer-Larsen \& 
Dolgov 1999; Hogan 1999) is designed to aid problems 7 \& 9 in two ways: WDM 
particles stream freely in the early universe suppressing small-scale power.  
They also have larger velocity spreads in halos of greater volume density, 
because of Liouville's theorem, thereby precluding strong density gradients.  
However, fermionic WDM falls foul of problem 6, while bosonic warm DM, which has 
a small fraction of very high phase space density material (\eg\ Madsen 2000), 
would probably again form cusped halos as in CDM.  Self-interacting dark matter 
(Spergel \& Steinhardt 2000) can in principle solve problems 7 \& 9 (Burkert 
2000; Dav\'e \etal\ 2000) but only if the cross section is tuned; it may also 
have difficulties with its prediction of spherical halos.  We emphasize that 
hypothetical modifications to the properties of DM particles have been motivated 
mostly by problems 7 \& 9, and any collisional effect removes 6, but none 
address the much more difficult problems 1--5.

\section{The Alternative}
If the problems listed in \S3 cannot be solved in the context of an acceptably 
simple DM model, the only logical alternative is to modify gravity in the 
weak-field regime.  No satisfactory theory that achieves this has yet been 
proposed.

Remarkably, we already know the effective force-law for galaxies (McGaugh 
1999a): MOND (Milgrom 1983) is a highly successful phenomenological description 
of the dynamical data on spiral galaxies which effortlessly resolves problems 1 
through 4, 6, 10 \& 11 above.  Thus it disposes of those most intractable in DM 
models (although problem 3 is ``resolved'' only by introducing a new fundamental 
physical scale).  Problem 2 was a notable prediction of MOND.  It seems not 
unreasonable to imagine that the other problems might also be resolved once a 
theory of galaxy formation is formulated in some modified gravity theory.

But modifying the law of gravitation is not a panacea; in particular it faces 
serious challenges in reproducing the successes of standard cosmology.  The 
following is a list of some more obvious difficulties.

\smallskip
\begin{enumerate}

\item A modified theory must be cast as a generally covariant theory to 
reproduce Einstein gravity in its well-tested regimes.  So far, attempts to fit 
MOND into a natural metric framework have been largely unsuccessful (Sanders 
1997).

\item It must establish a cosmology in which the scale factor evolution yields 
the first CMB acoustic peak at its observed angular scale.

\item Without DM, Silk damping (Silk 1968) at the time of recombination would 
eradicate primordial fluctuations on scales of galaxies and smaller.  A new 
source of fluctuations to seed the growth of galaxies would therefore be 
required (Sanders 1998).

\item In the DM picture, the tiny fluctuations observed in the microwave 
background have grown to the observed current large scale structure only because 
the DM density variations started growing before recombination.  A cosmology 
without DM requires other means to increase the growth factor since the time of 
recombination (Sanders 1998).

\item Observed gravitational lensing must arise solely from the baryonic matter 
distribution.

\end{enumerate}

\noindent
One other issue might be added to both this list of problems and that in \S3.  
Current observational data strongly indicate a cosmological constant (\eg\ Jaffe 
\etal\ 2000), while conventional physical theories offer no explanation for its 
tiny value (Weinberg 2000) and it requires us to live at a special time in the 
evolution of the universe.  It is conceivable that a dramatic alteration of 
gravitation and cosmology might provide a different reason for accelerated 
expansion.

Future signatures of the absence of DM might include a still more strongly 
suppressed third acoustic peak in the CMB.  In the absence of any driving force 
from dark matter, its amplitude should be further suppressed relative to the 
second peak due to Silk damping, while most conventional models predict an 
enhanced third peak from driven baryon compression.  Another possible test is 
the imprint of these peaks on the large-scale structure power spectrum 
(Eisenstein \& Hu 1998): without a dominant contribution of DM to the 
gravitational potential, substantial oscillations in the power spectrum of 
large-scale structure might survive.

\section{Conclusions}
Dark matter models are facing increasingly serious difficulties on galaxy scales 
as the quality of both the simulations and the data constraining the properties 
of galaxy halos accumulates.  Those who suggest that a revised law of gravity in 
the ultra-weak field limit might be more natural (Milgrom 1983; Sanders 1997; 
McGaugh \& de Blok 1998; Mannheim 2000; and others) are regarded as radicals.  
However, theories incorporating both a fine-tuned cosmological constant and dark 
matter particles with {\it ad hoc\/} properties have lost the compelling 
aesthetic simplicity of the original CDM model.

The problem of formulating a generally covariant, alternative theory of gravity 
upon which a competing model of structure formation can be built is itself a 
severe challenge.  Without such a theory, we cannot say whether the other 
problems on our preliminary list in \S5 will turn out to be harder to resolve 
than those of \S3, and it is possible that the list would grow to become still 
more daunting if such a theory could be developed.  But the difficulties facing 
DM models and the success of the MOND phenomenology suggest that a serious 
examination of alternative gravity solutions is called for. 

\acknowledgements We thank Paul Steinhardt and Michael Turner for critical 
readings of a draft version.  Comments from Stacy McGaugh, Bob Sanders, David 
Spergel, Matthias Steinmetz and Scott Tremaine were also helpful.  This work was 
supported by NASA LTSA grant NAG 5-6037 and NASA ATP grant NAG 5-7015.  AK is a 
Cotrell Scholar of the Research Corporation.

\end{document}